\documentclass[pre,amsmath,amssymb,twocolumn,showpacs,amsfonts,twocolumn]{revtex4}
\usepackage{graphicx}

\def \beq {\begin{equation}}
\def \eeq {\end{equation}}
\def \tr {\rm Tr}

\def \beq {\begin{equation}}
\def \eeq {\end{equation}}

\begin{document}
\title{The Quantum Zeno Effect Immunizes the Avian Compass Against the Deleterious Effects of Exchange and Dipolar Interactions}
\author{A. T. Dellis and I. K. Kominis}
\email{ikominis@iesl.forth.gr}

\affiliation{Department of Physics, University of Crete, Heraklion
71103, Greece}

\date{\today}

\begin{abstract}
Magnetic-sensitive radical-ion-pair reactions are understood to
underlie the biochemical magnetic compass used by avian species
for navigation. Recent experiments have provided growing evidence
for the radical-ion-pair magnetoreception mechanism, while recent
theoretical advances have unravelled the quantum nature of
radical-ion-pair reactions, which were shown to manifest a host of
quantum-information-science concepts and effects, like quantum measurement, quantum jumps and the
quantum Zeno effect. We here show that the quantum Zeno effect
provides for the robustness of the avian compass mechanism, and
immunizes it's magnetic and angular sensitivity against the
deleterious and molecule-specific exchange and dipolar
interactions.
\end{abstract}
\maketitle
\section{Introduction}
"In the history of natural selection, did nature ever come across
a way to use quantum weirdness?" This is a question claimed
\cite{lloyd} to have an affirmative answer, at least in regard
with the apparent ability of photosynthetic antennae to
efficiently guide the excitonic energy to the photosynthetic
reaction center. Questions like the previous one, addressing the
possibility of biological processes exhibiting non-trivial quantum
effects, ordinarily thought to be suppressed in the
decoherence-prone biological environment \cite{davies,abbott},
have attracted an increasing attention in recent years. For
example, significant experimental \cite{engel,lee} and theoretical
\cite{plenio,guzik} progress has been recently made on elucidating
the role of quantum coherence and quantum walks, respectively, in
the workings of photosynthetic antennae complexes. In a different
front, radical-ion-pair reactions \cite{schulten,steiner} have
been recently shown \cite{kom1,JH,komNJP,briegel,vedral,cai,kom2} to exhibit the full machinery of
concepts and physical effects familiar from quantum information
science. Radical-ion pairs play a fundamental role in a series of
biologically relevant chemical reactions, ranging from charge
transfer initiated reactions in photosynthetic reaction centers
\cite{boxer} to magnetic sensitive reactions abounding in the
field of spin-chemistry \cite{timmel}. In particular, radical-ion
pairs are understood to underlie the biochemical magnetic compass
used by avian species to navigate in earth's magnetic field
\cite{ritz,ww}, as corroborated by several recent experiments
\cite{ritz_res,maeda,ritz_hore,zapka}. 

\begin{figure}
\includegraphics[width=7.5 cm]{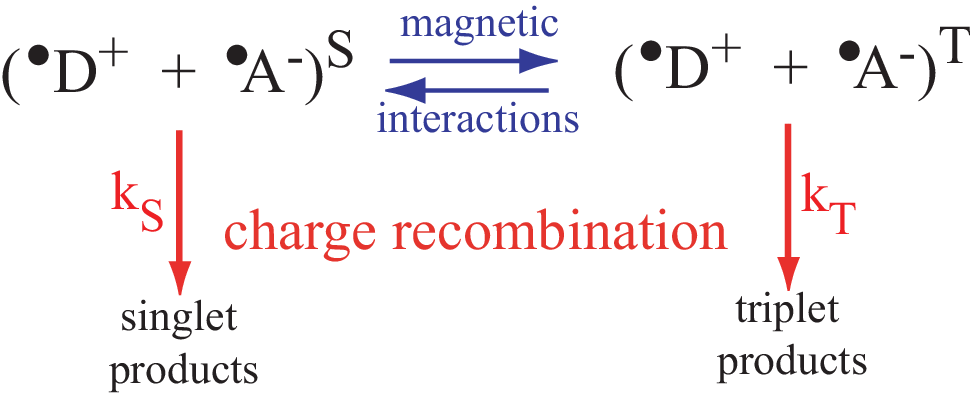}
\caption{Radical-ion-pair reaction dynamics: Photoexcitation of a
donor-acceptor molecule DA followed by charge-transfer creates a
radical-ion-pair, i.e. two molecular ions and two unpaired
electrons (two dots). The Zeeman interaction of the two unpaired
electrons with the external magnetic field and hyperfine
interactions with the molecule's magnetic nuclei induce a coherent
singlet-triplet conversion, ceased by the spin-selective charge
recombination, which transforms singlet (triplet) radical-ion
pairs into singlet (triplet) neutral products at a rate $k_S$
($k_T$).} \label{rmodel}
\end{figure}

In Fig. \ref{rmodel} we depict a generic model for radical-ion-pair
reactions, which form a magnetic sensor since the reaction product
yields depend on the external magnetic field. Radical-ion pairs
are formed by a charge transfer process following a
photoexcitation of a donor-acceptor dyad, leading to two molecular
ions and two unpaired electrons. The latter can either be in the
spin singlet or in the spin triplet state. Magnetic interactions
with the external magnetic field and hyperfine interactions with
the molecule's magnetic nuclei bring about a coherent
singlet-triplet oscillation. At some random instant in time the
reaction is terminated, since the radical-ion-pair undergoes
charge recombination, leading to the reaction products. Angular
momentum conservation enforces spin selectivity of the
recombination process, i.e. singlet (triplet) radical-ion pairs
recombine to singlet (triplet) neutral products. Moreover,
anisotropic hyperfine interactions within the molecule render the
reaction yields dependent on the inclination of the external
magnetic field with respect to a molecule-fixed coordinate frame.

Interestingly, intra-molecule magnetic interactions are more
complicated. Both spin-exchange and long-range dipolar
interactions affect reaction dynamics, to an extent dependent on
the particular molecular structure. In this respect, it has been
recently shown \cite{efimova} that the presence of exchange and/or
dipolar interactions significantly suppresses the magnetic and
angular sensitivity of the reaction yields, thus severely
degrading the mechanism's functionality. Along the same lines, it
was concluded \cite{efimova} that only when the molecular
parameters determining $J$ and $D$, the exchange and dipolar
couplings, are fine-tuned so that the effects of these two
interactions cancel each other, is the magnetic and angular
sensitivity of the reaction restored. Although it is conceivable that Nature has conjured up such a
fortuitous cancellation for a functionally important biological
sensor, the fact that $J$ and $D$ depend sensitively on
molecule-specific parameters, like the donor-acceptor distance $r$ (for example $J$ has
an exponential dependence on $r$) makes this possibility questionable.

It was recently shown \cite{kom1} that radical-ion-pair reactions
form a biochemical system that exhibits the quantum Zeno effect \cite{qZ}.
We will here show that when the quantum Zeno effect is manifested (i.e. when the 
recombination rates are asymmetric), the reaction's magnetic and angular sensitivity is practically
independent of the presence or not of exchange and/or dipolar
interactions. This realization has profound implications for the
robustness of this biological sensor, i.e. a non-trivial quantum
effect renders the sensor insensitive to molecule-specific
parameters \cite{efimova}, such as donor-acceptor distance
(affecting the exchange coupling and the long-range dipolar
coupling) and the inter-radical medium and the particular
electronic structure (affecting the exchange coupling). Whether
Nature has engineered molecules realizing a fine-tuned
cancellation \cite{efimova} of the adverse effects of exchange and
dipolar interactions, or on the other hand, has evolved the avian
compass into operating at the quantum Zeno regime, remains to be
discovered. In the following we will analyze the merits of the
latter possibility. It is noted that the topic of this work is 
at the center of two debates. The first has to do with whether the 
avian magnetoreception is based on the radical-pair mechanism or magnetic
nano-particles, as suggested by several authors \cite{mn}.
We do not make any suggestion as to which mechanism is actually responsible for 
avian magnetic navigation. We just deal with a particular weakness of the radical-pair magnetoreception as 
described in \cite{efimova} and suggest how the radical-pair magnetoreception can be indeed viable 
in the appropriate parameter regime. The second debate has to do with the fundamental master
equation describing radical-ion-pair reactions \cite{kom1,JH,comment,reply,kom2,JH2,ivanov}. In particular, 
there are currently three different theories describing the quantum dynamics of these reactions, the traditional theory (also referred to as Haberkorn master equation), the 
Jones-Hore theory and the theory developed by one of us.  
As will be shown in the following, the results of this work are qualitatively valid for all three theories, the only differences
being quantitative.

In Section II we reiterate the quantum dynamics of
radical-ion-pair reactions and elaborate on the magnetic
interactions within the radical-ion-pair central to the problem of
study. In Section III we analyze the magnetic and angular
precision of the avian compass magnetic sensor in the presence of
exchange interactions, while in Section IV we explain the robust
performance of the avian compass as a direct consequence of the
quantum Zeno effect and the spin delocalization resulting from the
quantum measurement dynamics inherent in radical-ion-pair
recombination reactions.
\section{Quantum Dynamics and Magnetic Interactions in the Radical-Ion-Pair Avian Compass}
What is of interest in describing radical-ion-pair reactions is
the spin state of the pertaining particles, the two electrons and
the molecule's nuclear spins. The spin state of the
radical-ion-pair is described by a $4n$-dimensional density matrix
$\rho$ , where the factor 4 is the spin multiplicity of the two
electrons and $n=(2I_{1}+1)(2I_{2}+1)...(2I_{k}+1)$ is the nuclear
spin multiplicity of the molecule's $k$ nuclei having nuclear
spins $I_1$, $I_2$,...,$I_k$. The time evolution of $\rho$
is described by a master equation of the form 
\beq
d\rho/dt=-i[{\cal H}_{m},\rho]-{\cal L}(\rho)\label{qev}
\eeq
where the first term describes the unitary evolution of $\rho$ due to the
magnetic interactions embodied in ${\cal H}_{m}$ and ${\cal L}$ denotes a superoperator
that takes into account the reaction dynamics. It is this part of the theoretical description that three above mentioned
theories differ, and the relevant details can be found in the recent literature \cite{kom1,JH,kom2}.
The two basic parameters and operators that determine the reaction dynamics are
the singlet and triplet recombination rates $k_S$ and $k_T$, and the singlet and triplet projection operators, $Q_S$ and $Q_T$,
respectively. Once the density matrix evolution is known, i.e. once the master equation \eqref{qev} is solved, the 
reaction yield e.g. the triplet can be calculated from
\beq
Y_{T}=k_{T}\int_{0}^{\infty}\tr\{\rho Q_{T}\}\label{tyield}
\eeq
The magnetic Hamiltonian for the problem under study, ${\cal H}_{m}={\cal H}_{Z}+{\cal
H}_{hf}+{\cal H}_{ex}$, is composed of ${\cal H}_{Z}$, the Zeeman
interaction of the two unpaired electrons (nuclear Zeeman
interaction is negligible) with the external magnetic field,
${\cal H}_{hf}$, the hyperfine couplings of the electrons with the
surrounding nuclear spins, and finally the spin-exchange
interaction, ${\cal H}_{ex}$. For the transparency of the
following discussion we will ignore the dipolar interaction as its
inclusion leads to exactly the same conclusions. The Zeeman
interaction Hamiltonian that will be used for the study of the
magnetic sensitivity is ${\cal H}_{Z,magn}=\omega(s_{1z}+s_{2z})$,
where the magnetic field of magnitude $B$ is assumed to be in the
$z$-axis ($\omega=\gamma B$, with $\gamma=2\pi\times 2.8  {\rm MHz/G}$). For the study
of the angular sensitivity we take the magnetic field, again of
magnitude $B$, to be in the x-y plane, hence the Zeeman
interaction term will be ${\cal
H}_{Z,ang}=\omega\cos\phi(s_{1x}+s_{2x})+\omega\sin\phi(s_{1y}+s_{2y})$. For the study of the magnetic sensitivity we 
vary $B$, whereas for studying anfular sensitivity we keep $B$ constant and vary the angle $\phi$.
In the following we will consider the simplest physically
realizable radical-ion-pair containing just one spin-1/2 nucleus
(in which case dim($\rho$)=8), hence the hyperfine interaction
Hamiltonian is ${\cal
H}_{hf}=\mathbf{I}\cdot\mathbf{A}\cdot\mathbf{s}_{1}$, where
$\mathbf{A}$ is the hyperfine coupling tensor of the single
nuclear spin $\mathbf{I}$ existing in e.g. the donor molecule with
the donor's unpaired electron. We will consider the simplest case
where the hyperfine tensor is diagonal with one non-zero component
$A_{xx}=a$, to provide for the angular sensitivity on the $x-y$
plane. Thus ${\cal H}_{hf}=as_{1x}I_{x}$. Finally the
spin-exchange Hamiltonian is ${\cal
H}_{ex}=J\mathbf{s}_{1}\cdot\mathbf{s}_{2}$. 
We note that the simplification of considering just one nuclear spin is common in all such
considerations, and although it does not exhaust all the richness of phenomena that can be 
observed by the realistic inclusion of more nuclear spins (as is the case in Nature), it does
provide an idea of what is in principle feasible, and this is exactly the goal of this work. 

In the following we will calculate the magnetic and angular sensitivity of the
reaction for two regimes: (i) the "traditional" regime with equal 
recombination rates $k_S=k_{T}$ on the order of or smaller than the hyperfine coupling $a$. It is in this regime that almost
all calculations have been performed based on the previous,
traditional master equation. We then study the regime (ii)
where $k_{T}\gg k_{S}$ with $k_{T}$ on the order of or larger than the hyperfine coupling $a$, i.e. when the
quantum Zeno effect is manifested. To elaborate on this, we note that if the initial state
of the molecule is the singlet (which is usually the case) and there
exist asymmetric recombination rates then the spin
state of the radical-pair is strongly projected to the singlet state by
the triplet reservoir. As has been explained in \cite{kom1,kom2}, the 
singlet and triplet reservoirs essentially measure the observable $Q_{S}$ at
a total measurement rate of $(k_{S}+k_{T})/2$. A large measurement rate essentially
means frequent quantum jumps to either the singlet or the triplet state. Since the molecule
starts out from the singlet, chances are that most of those jumps will be to the singlet state, hence
the strong projection to the singlet, which is the signature of the quantum Zeno effect, or in other words,
the strong measurement regime (the same considerations obviously apply to the case of a triplet initial state and $k_{S}\gg k_{T}$). We will then show that in the regime (i) the inclusion of the spin exchange interaction
indeed degrades the magnetic and angular sensitivity of the
reaction, as has already been analyzed \cite{efimova}. However,
regime (ii) exhibits an appreciable magnetic and angular
sensitivity with their dependence on the exchange coupling $J$
being significantly suppressed. 
\section{Magnetic and Angular Sensitivity of the Avian Compass}
In Figures 2a and 2b we plot an example of the triplet reaction yield $Y_T$, calculated from \eqref{tyield},
as a function of the external magnetic field (using ${\cal
H}_{Z,magn}$) and the field's angle (using ${\cal H}_{Z,ang}$), respectively.
The magnetic sensitivity of the reaction at earth's field of interest for the avian compass 
is proportional to the slope of $Y_T$ vs $B$ 
calculated at $B$=0.5 G. Similarly, the angular sensitivity of the reaction is proportional to the (maximum) slope of $Y_T$ vs $\phi$. The smallest measurable change of the magnetic field, $\delta B$ (absolute magnetic sensitivity), and the smallest detectable change, $\delta\phi$, in the field's angle with respect to the molecule's $x$-axis (absolute angular sensitivity or heading
error) both follow from the previous calculations if the smallest
measurable reaction yield change, $\delta Y_{T}$, is known. It
thus follows that
\begin{align}
\delta B&={{\delta Y_{T}}\over {|dY_{T}/dB|_{B=0.5~{\rm G}}}}\label{deltaB}\\
\delta\phi &={{\delta Y_{T}}\over {(Y_{T,max}-Y_{T,min})/\Delta\phi}}\label{deltaphi}
\end{align}
where $Y_{T,min}$ and $Y_{T,max}$  are the minimum and maximum
values of the yield $Y_{T}(\phi)$ and $\Delta\phi=90^{\rm o}$  is
the angular width of the full swing between $Y_{T,min}$ and
$Y_{T,max}$.
\begin{figure}
\includegraphics[width=8.5 cm]{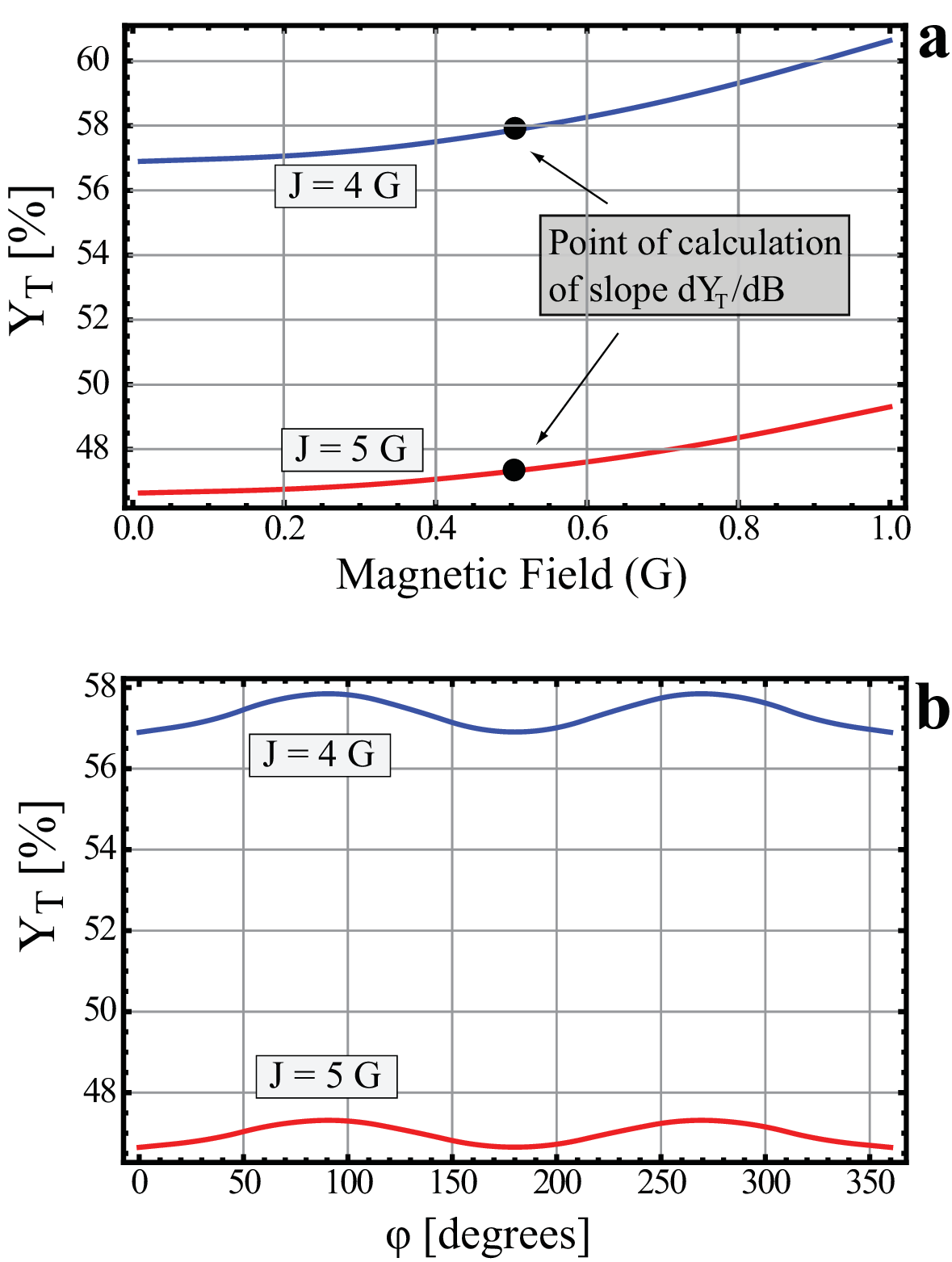}
\caption{Examples of (a) magnetic and (b) angular sensitivity of the triplet reaction yield as a function of (a) magnetic field and (b) magnetic field's direction in the horizontal plane for two different values of the 
exchange coupling $J$. The calculations were done for $k_S$=0.5 MHz, $k_T$=40 MHz, hyperfine coupling $a$=1.75 G and a magnetic field $B=0.5$ G for the angular sensitivity.} \label{f1}
\end{figure}
To make further progress the value of $\delta Y_{T}$ must be known or estimated.
Obviously $\delta Y_{T}$ depends on the particular realization of
the biochemical mechanism transducing the radical-ion-pair
reaction yield to a physiological signal. On rather general
grounds it has been shown \cite{weaver} that $\delta Y_{T}$ is
connected to $N_R$, the number of neuronal receptors sensitive to
the radical-ion-pair reaction product molecules, by $(\delta
Y_{T})^{2}=4/N_{R}$. We choose $N_{R}=1.6\times 10^{7}$, in order
to set $\delta Y_{T}$ at the value $\delta Y_{T}=0.05\%$. The
chosen value of $N_R$ and hence $\delta Y_{T}$ is realistic
\cite{weaver} and has the consequence that it sets the magnetic
sensitivity at zero exchange coupling at the
value of $\delta B\approx 0.01$ G, i.e. at 2\% of earth's field.
This level of magnetic sensitivity is understood \cite{ww} to be actually
realized in several avian species. It is stressed, however, that
the following considerations are qualitatively independent of the
particular value of $\delta Y_{T}$, which just sets the absolute
scale of the derived magnetic and angular sensitivity. 

From plots like the ones in Figs. 2a and 2b, and for various values of the 
exchange coupling $J$, we obtain the sensitivities $\delta B$ and $\delta\phi$, according to
\eqref{deltaB} and \eqref{deltaphi}, which are plotted in Figs. 3a and 3b, respectively.
It is clearly seen in Figure 3a that in the traditional regime (i) the
magnetic precision plunges to $\delta B=0.5$ G already at $J\approx$ 6 G. 
\begin{figure}
\includegraphics[width=8.5 cm]{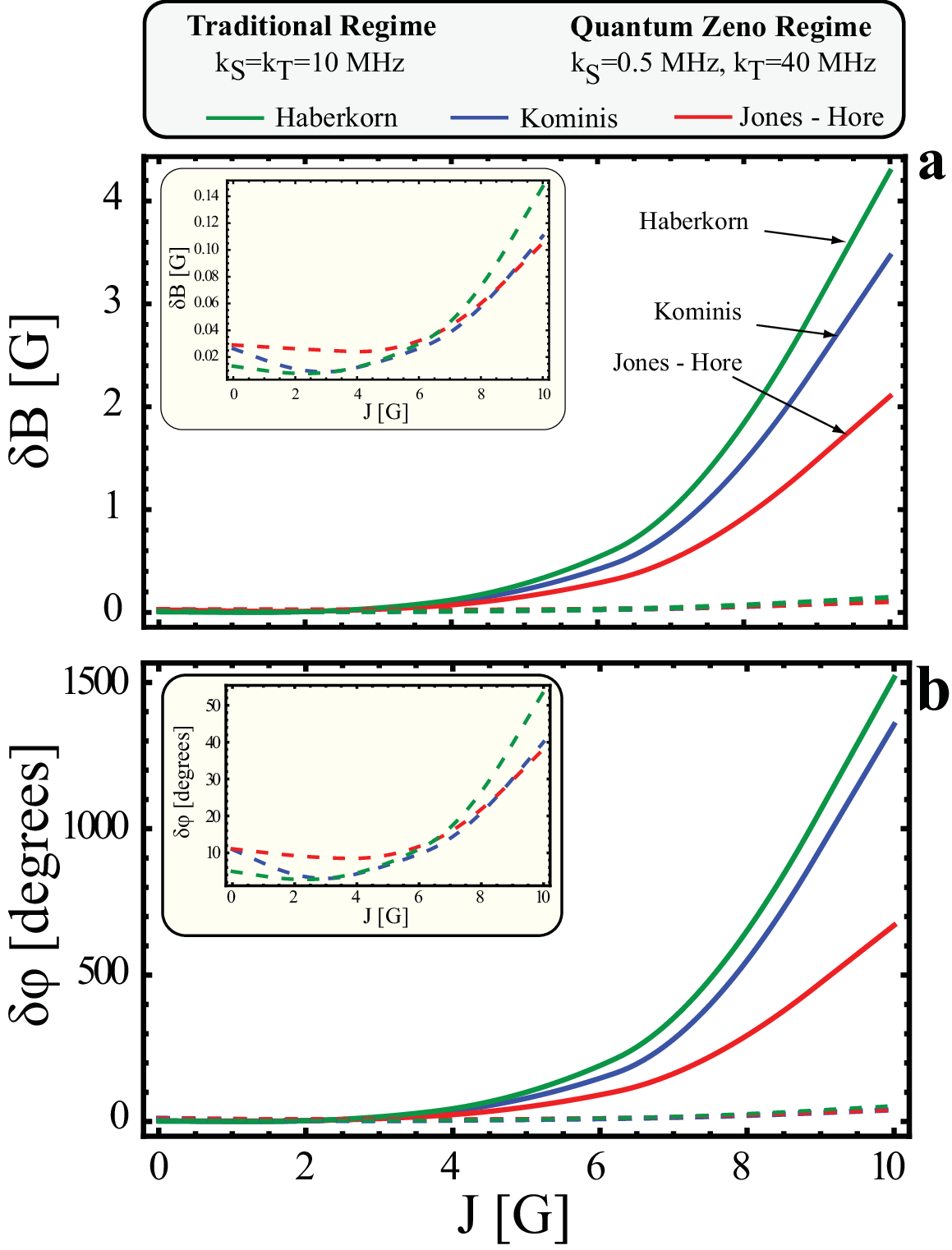}
\caption{(a) Magnetic sensitivity and (b) angular sensitivity (heading error) of the reaction as a function of the 
exchange couplijng $J$ for the two different regimes of (i) equal recombination rates (solid lines) and (ii) asymmetric recombination rates (dashed lines) and for all three theories. Insets zoom into low $\delta B$ and $\delta\phi$ values in order to make the predictions of all three 
theories in regime (ii) distinguishable. For all calculations $k_S$=$k_T$=10 MHz for the traditional regime, $k_S$=0.5 MHz and $k_T$=40 MHz for the Zeno regime and $a$=1.75 G. For the angular sensitivity calculation the magnitude of the magnetic field was $B=0.5$ G. It is obvious that both the magnetic and the angular sensitivity depend on $J$ much less 
sensitively in regime (ii), where the strong projective measurement induced by a large $k_T$ (quantum Zeno effect) dominates the dynamics.} \label{f2}
\end{figure}
Similarly, as shown in Figure 3b, the angular precision in regime
(i) drops dramatically with increasing $J$, with a complete loss
of heading information already at $J$=6 G. In contrast, in the
quantum Zeno regime (ii) the angular precision of about
$\delta\phi=40^{\rm o}$ at the highest value of the exchange
coupling is actually at the level of experimental
observations \cite{headingErr} of the heading error of the avian compass.
 Finally, as noted before, all three 
theories produce qualitatively similar results. The particular values obtained here for the absolute sensitivities $\delta B$ and $\delta\phi$ obviously depend on the particular hyperfine couplings used and the chosen values of the recombination rates. 
As pointed out in the introduction, more complicated models will result in different numbers, however, our sole goal is to demonstrate a behaviour that is in principle feasible.
\section{Explanation of the Robust Avian Compass Sensitivity}
We will now explain the robust magnetic and angular sensitivity resulting in the quantum Zeno regime. This follows
by considering the behavior of unrecombined radical-ion pairs, described by the maser equation
\beq
d\rho/dt=-i[{\cal H}_{m},\rho]-(k_{S}+k_{T})(Q_{S}\rho+\rho Q_{S}-2Q_{S}\rho Q_{S})/2\label{MEnr}
\eeq
This master equation has been derived in \cite{kom1} and its physical meaning explained in detail
in \cite{kom2}. Essentially, unrecombined radical-ion pairs suffer a loss of singlet-triplet coherence due to
the continuous measurement of $Q_{S}$ induced by the singlet and triplet reservoirs and the concomitant quantum jumps. 
To get an insight into the dynamics in the regime of the asymmetric recombination rates, we consider the eigenvalues of the master
equation \eqref{MEnr}, which are obtained by diagonalizing the
matrix ${\cal M}$ (of dimension ${\rm dim}(\rho)^{2}$) that
satisfies $d\tilde{\rho}/dt={\cal M}\tilde{\rho}$, where
$\tilde{\rho}$ is a column matrix containing all matrix elements
of $\rho$ . The resulting eigenvalues are of the form
$-\lambda+i\Omega$, with $\lambda\geq 0$ being the decay rate and
$\Omega$ the oscillation frequency of the particular eigenmode. As
is in general the case with the quantum Zeno effect \cite{streed},
some of the eigenvalues have decay rates increasing with the
measurement rate $k$ as $\lambda\sim k$, while the others
(responsible for the quantum Zeno effect) decrease with $k$ as
$\lambda_{qZ}\sim h^{2}/k$, where $h$ is the characteristic
frequency scale of the system, here determined by the magnetic
Hamiltonian ${\cal H}_{m}$. In our case, the measurement rate $k=(k_{S}+k_{T})/2$, and in the
quantum Zeno regime in which $k_{T}\gg k_{S}$, it will be $k\approx k_{T}/2$.

Now, the exchange Hamiltonian can be written (up to an additive
constant) as ${\cal H}_{ex}=-JQ_{S}$. Furthermore, as is known
from quantum measurement theory \cite{bp}, the deterministic
evolution of the system's quantum state (due to the unitary
Hamiltonian evolution and the measurement of $Q_S$ with rate $k$)
is generated by the non-hermitian operator ${\cal K}={\cal
H}_{m}-ikQ_{S}$. It is easily seen that if ${\cal H}_{m}^{J=0}$ is
the magnetic Hamiltonian without the exchange interaction ${\cal
H}_{ex}$, then ${\cal K}={\cal H}_{m}^{J=0}-i(k-iJ)Q_{S}$, i.e.
the inclusion of the exchange interaction is equivalent to
replacing $k$ with with an imaginary measurement rate $k-iJ$. We
can now complete the argument as follows: the eigenvalues with a
real part that scales as $\lambda\sim k$ pick up an oscillation
frequency (in addition to $\Omega$) of $-J$, the effect of which
roughly averages out. On the other hand, the eigenvalues with the
quantum Zeno scaling $\lambda_{qZ}\sim h^{2}/k$ suffer a change in
their real part which becomes (since in our case $k/J< 1$)
$\lambda_{qZ}'\approx h^{2}k/J^{2}\ll \lambda_{qZ}$. Thus, with increasing $J$, the spin
state evolution is slowed down. This can be clearly seen in Figure 4a, which shows the time evolution of the normalization of the density matrix, $\tr\{\rho\}$, i.e. the number of existing radical-ion
pairs, as calculated from \eqref{qev}. The reaction is considered to be terminated
when $\tr\{\rho\}\approx 5\times 10^{-4}$, i.e. when the reaction
yield is known to within $\delta Y_{T}$. It is clearly seen that
in the quantum Zeno regime, the reaction time depends on $J$ in
the way outlined before.
\begin{figure}
\includegraphics[width=8.30 cm]{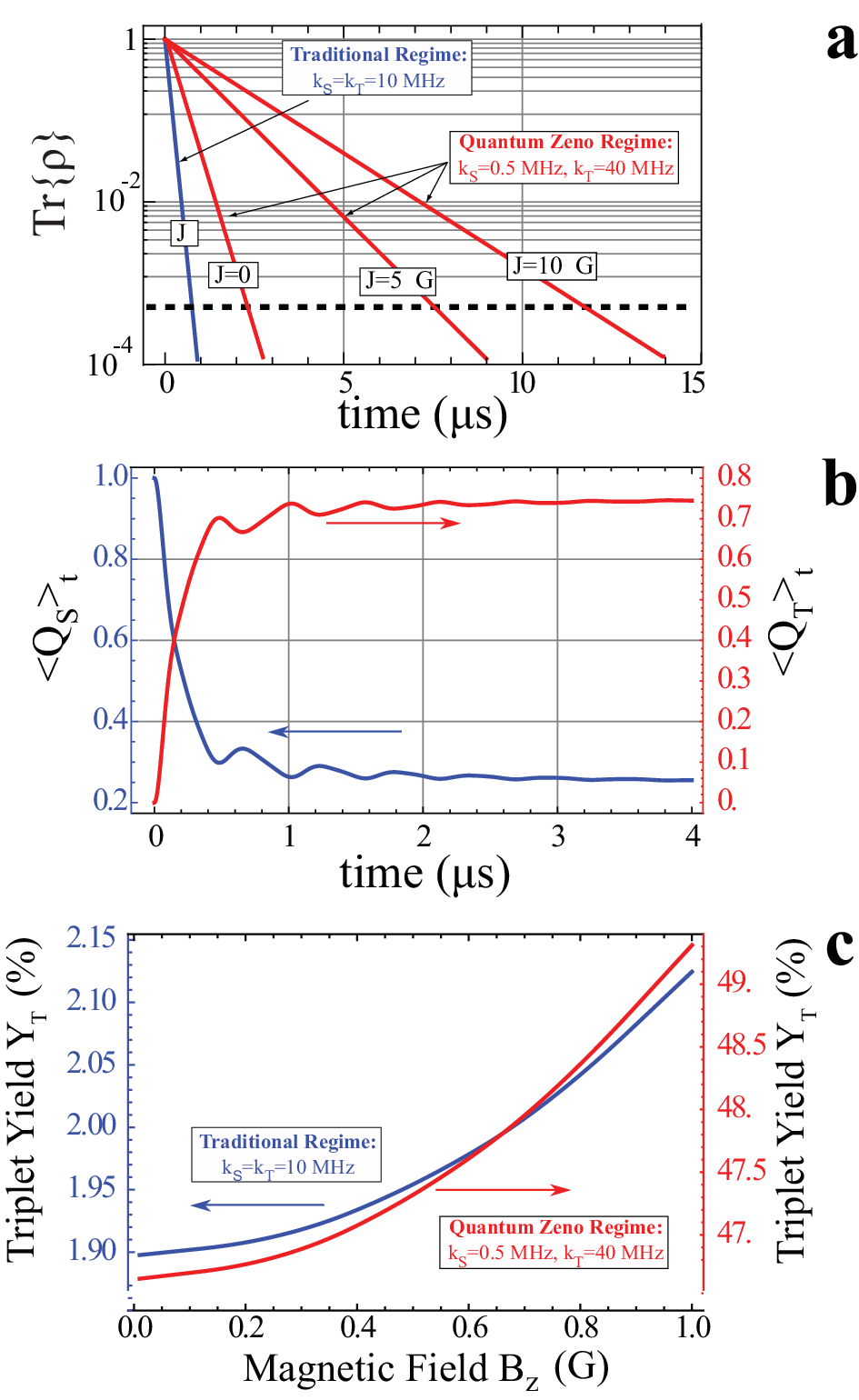}
\caption{(a) Time evolution of the normalization of $\rho$. The dashed line at $N=5\times
10^{-4}$ signifies the "termination" of the reaction, i.e. the
point when the remaining radical-ion pairs are 0.05\% of the
initial number. It is seen that in the quantum
Zeno regime, the reaction lasts longer for increasing $J$, as
explained in the text. In contrast, in the traditional regime of
equal recombination rates the duration of the reaction is
independent of $J$. (b) Time evolution of the singlet and triplet
probability, $\langle Q_{S}(t)\rangle$ and $\langle
Q_{T}(t)\rangle$, respectively for the unrecombined radical-pairs, calculated from \eqref{MEnr} for $B$=0.5 G and $J$=10 G. The
measurement dynamics inherent in the charge recombination process
of radical-ion pairs "delocalize" the electron spin state at long
times. (c) Triplet yield as a function of the magnetic field for
$J$=10 G, plotted in the two regimes for the recombination rates.
} \label{f3}
\end{figure}
On the contrary, when $k_{S}=k_{T}=\kappa$, the change of $\tr\{\rho\}$ during the time interval $dt$ easily follows from \eqref{qev} 
and is $d\tr\{\rho\}=-dt(k_{S}\langle
Q_{S}\rangle+k_{T}\langle Q_{T}\rangle)=-\kappa dt\tr\{\rho\}$, since $Q_{S}+Q_{T}=1$. Hence
 the reaction time is proportional
to $1/\kappa$ and independent of $J$. The result is that during
the short reaction time in the traditional regime, the triplet
probability $Q_{T}$ has not increased appreciably (Figure 4b), and
the triplet yield is small, as shown in Figure 4c. In contrast, in
the quantum Zeno regime the reaction has enough time to "sample"
large values of $Q_{T}$ and lead to a triplet yield about an order
of magnitude higher, hence the higher sensitivity in this regime.
In other words, as seen in Figure 4c, the relative change $\delta
Y_{T}/Y_{T}$ of the triplet yield with the magnetic field is
roughly the same in both cases, but the absolute value of $Y_T$ differs
by a factor of 20, leading to respectively high slopes $dY_{T}/dB$
and $dY_{T}/d\phi$. To summarize, the quantum
measurement dynamics inherent in the recombination process of
radical-ion pairs result in "delocalization" of the electron spin
state at long times, as evidenced in Figure 4b. The asymmetric
($k_{T}\gg k_{S}$) recombination rates result in the $J$-dependence of
the reaction time. The interplay of these two effects provides
for the robust magnetic and angular sensitivity in the presence of
the exchange interaction. 
\subsection{Quantum Zeno Effect in the Traditional Master Equation}
The quantum Zeno effect is embodied also in the traditional theory as well as the Jones-Hore theory. This has been mentioned in \cite{ivanov} and 
analyzed in detail in \cite{berdinskii}. We will here elucidate this using a 
simple two-dimensional example. Consider the density matrix $\rho=\begin{pmatrix}\rho_{SS} & \rho_{ST}\\\rho_{TS} & \rho_{TT}\end{pmatrix}$ and a magnetic Hamiltonian of the form ${\cal H}=\begin{pmatrix}0 & \omega\\\omega & 0\end{pmatrix}$. The projection operators are in this case $Q_{S}=\begin{pmatrix}1 & 0\\0 & 0\end{pmatrix}$ and $Q_{T}=\begin{pmatrix}0 & 0\\0 & 1\end{pmatrix}$. The traditional master equation, $d\rho/dt=-i[{\cal H},\rho]-k_{S}(Q_{S}\rho+\rho Q_{S})/2-k_{T}(Q_{T}\rho+\rho Q_{T})/2$ is, assuming for simplicity that $k_{S}=0$, equivalent to 
\beq
{d\over {dt}}\begin{pmatrix}\rho_{SS}\\\rho_{ST}\\\rho_{TS}\\\rho_{TT}\end{pmatrix}=
\begin{pmatrix}0& i\omega & -i\omega &0\\
i\omega & -k_{T}/2 & 0 & -i\omega\\
-i\omega & 0 & -k_{T}/2 & i\omega\\
0 & -i\omega & i\omega & -k_{T}\end{pmatrix}\begin{pmatrix}\rho_{SS}\\\rho_{ST}\\\rho_{TS}\\\rho_{TT}\end{pmatrix}
\eeq
The above 4$\times 4$ matrix has four eigenvalues, $-k_{T}/2$ (doubly degenerate), $-k_{T}/2-\sqrt{k_{T}^{2}-16\omega^{2}}/2$ and 
$-k_{T}/2+\sqrt{k_{T}^{2}-16\omega^{2}}/2$. For $k_{T}\gg\omega$, the last eigenvalue is approximately equal to $-2\omega^{2}/k_{T}$. This, as already noted, is the quantum Zeno scaling, i.e. the larger the interrogation rate $k_{T}$, the slower the decay of the density matrix elements dependent on the particular eigenvalue. In other words, even if the traditional theory is not constructed on the quantum measurement concepts on which our theory is based, being a successful phenomenological theory it does bear part of the physics entering radical-ion-pair reactions in the asymmetric recombination regime. 
\section{Conclusions}
In conclusion, we have identified a concrete biological process in
which fundamental quantum effects have a profound effect on the
system's performance, alluding to the possibility that this
biological quantum sensor has evolved to a robust device by taking
advantage of non-trivial aspects of quantum physics.
Coincidentally or not, it turns out \cite{daviso} that the
radical-ion pairs participating in the last stages of the
electron-transfer processes taking place in bacterial
photosynthetic reaction centers operate at the quantum Zeno
regime, i.e. the triplet recombination rate $k_T$ is about 20
times larger than $k_S$, the singlet recombination rate. It is
noted that the manifestation of the quantum Zeno effect does not
require any parameter fine-tuning, but just the presence of
asymmetric recombination rates. This regime seems to offer an operational
advantage and hence the possibility that it is Nature's inevitable
choice is rather plausible. 

\section{Acknowledgements}
We would like to acknowledge the John S. Latsis Public Benefit Foundation for financial support under the 2011 scientific program project "Quantum-Limited Biochemical Magnetometers".

\end{document}